\documentclass{article}
\usepackage{graphicx}

\usepackage{hyperref}
\usepackage{amsmath}
\usepackage{amssymb}	
\usepackage{mathtools}

\usepackage{subcaption}
\usepackage[dvipsnames]{xcolor}

\usepackage{multirow}
\usepackage{pdfpages}
\usepackage{stfloats}
\usepackage{tabularx}
\usepackage[export]{adjustbox}
\usepackage{float}
\usepackage[
singlelinecheck=false 
]{caption}

\usepackage{multicol}

\begin{document}
\title{\bf Galactic black hole immersed in a dark halo with its surrounding thin accretion disk}
\author{{ Mohaddese Heydari-Fard$^{1}$ \thanks{Electronic address: m\_heydarifard@sbu.ac.ir}, Malihe Heydari-Fard$^{2}$\thanks{Electronic address:heydarifard@qom.ac.ir } and Nematollah Riazi$^{1}$\thanks{Electronic address: n\_riazi@sbu.ac.ir}}\\{\small \emph{$^{1}$ Department of Physics, Shahid Beheshti University, G. C., Evin, Tehran 19839, Iran}}\\{\small \emph{$^{2}$ Department of Physics, The University of Qom, 3716146611, Qom, Iran}}}
\maketitle

\begin{abstract}
By considering the analytic, static and spherically symmetric solution for the Schwarzschild black holes immersed in dark matter fluid with non-zero tangential pressure \cite{Jusufi:2022jxu} and Hernquist-type density profiles \cite{Cardoso}, we compute the luminosity of accretion disk. We study the circular motion of test particles in accretion disk and calculate the radius of the innermost stable circular orbits. Using the steady-state Novikov-Thorne model we also compute the observational characteristics of such black hole's accretion disk and compare our results with the usual Schwarzschild black hole in the absence of dark matter fluid. We find that the tangential pressure plays a significant role in decreasing the size of the innermost stable circular orbits and thus increases the luminosity of black hole's accretion disk.
\vspace{5mm}
\vspace{1mm}
\end{abstract}

\begin{multicols}{2}
\section{Introduction}
According to the standard model, the Universe mostly is composed of dark energy (68\%), dark matter (27\%) and baryonic matter which its contribution is only 5\% of the total mass of the Universe \cite{Planck:2015fie}. Observational evidence including the rotation curve of spiral galaxies \cite{Rubin}, dynamic of galaxy clusters \cite{Zwicky}, the cosmic microwave background radiation \cite{Planck}, extremely high mass-luminosity ratios of elliptical galaxies, etc., indicate that dark matter is clustered in the center of galaxies. On the other hand, it is widely established that the center of many galaxies contains supermassive black holes (BHs) \cite {Melia:2001dy, Genzel:2010zy}; so BHs in the center of galaxies have been immersed in dark matter halo.

Today BHs not only play an important role for testing general relativity (GR) in strong gravity regime but also are significant tools for testing the modified gravity theories. In recent years, due to the astronomical observations such as the orbital motion of the S-star in the center of our Milky Way galaxy \cite{Gillessen:2017jxc}--\cite{Gravity:2019nxk}, the gravitational waves detected by LIGO and Virgo detectors originate from crashing together and merging of two massive  BHs \cite{LIGOScientific:2016aoc}, the shadow image of M87* BH in center of M87 galaxy released in 2019 \cite{A1}--\cite{A6} and also Sgr A* BH in the center of our galaxy, released in May 2022 \cite{Sgr} by the Event Horizon Telescope (EHT) collaboration, the existence of BHs is widely established \cite{LIGOScientific:2016emj}--\cite{Antonucci:1993sg}.

Since the supermassive BHs at the center of galaxies are surrounded by dark matter halo, one expects that the dark matter distribution affects the geometry structure of BHs. Due to the density profile of dark matter halo around BH, the BH space-time is modified by dark halo \cite{Xu:2018wow}--\cite{Sadeghian:2013laa}. Recently, Cardoso et. al by solving Einstein equations with assuming anisotropic matter derived a galactic BH immersed in a dark halo with Hernquist-type density distribution, which is described by two free parameters $M$ and $a_{0}$ \cite {Cardoso}. The authors in Ref. \cite{Myung:2024tkz}, using the M87* and Sgr A* shadow data of the EHT collaboration, constrained two parameters $(M, a_{0})$ of dark matter halo. The shadow cast and gravitational lensing of such BHs were studied in Ref. \cite{Xavier:2023exm}. The shadow cast in a variety of background space-times have been also discussed in \cite{Heydari-Fard:2023ent}--\cite{MOG}. Also, for study about optical appearance of BHs surrounded by a dark matter halo and the tidal effects of a dark matter halo around a galactic BH, see Refs. \cite{Macedo:2024qky}--\cite{Liu:2022lrg} respectively. The galactic BH solutions surrounded by dark matter halo with different density profiles have been also obtained in \cite{Konoplya:2021ube}--\cite{Jha:2024ltc}. Similar to Cardoso's paper, the author in Ref. \cite {Jusufi:2022jxu} has constructed a new class of Einstein equation with anisotropic dark matter fluid with equation of state $P_{t}= \omega \rho$. Jusufi considered an object with mass $M$ in the center of galaxy immersed in anisotropic dark matter fluid and showed that unlike Cardoso’s paper, depending on the choice of mass profile one can obtain a central asymptotically flat BH and a naked singularity.

In addition to the observation of the BH shadow, the light emission of accretion disk also provides new important information about supermassive BHs \cite{Bambi}. The study of accretion disk process around a BH is another tool that could help to distinguish general relativity from modified gravity theories \cite{Moffat}. The amount of radiation energy and the spectrum emitted by the accretion flow is associated to geodesic motions of test particles and the geometry structure of the central compact body (BH). The simplest model of accretion disk process described by geometrically thin and optically thick disk model. Newtonian approach of such model first proposed by Shakura and Sunyaev \cite{Shakura} in 1973 and then Novikov and Thorne \cite{Novikov} extended it to the GR case. The observational properties of thin accretion disk around BHs in alternative gravity theories have been extensively investigated \cite{FR1}--\cite{PFDM}. The goal of this paper is to investigate the effects of dark matter halo and galaxy lengthscale on gaseous accretion disk surrounding supermassive BHs immersed in a dark halo proposed by Cardoso et al. \cite{Cardoso}.

The paper is organized as follows: In Section~\ref{2-BH}, we review the galactic BH solution surrounded by dark matter halo with Hernquist density distribution \cite{Cardoso}. The geodesic motion of test particles in the space–time of such BHs is studied in Section~\ref{3-geodesic}. In Section~\ref{4-disk}, we investigate the effect of dark matter halo and galaxy lengthscale on the electromagnetic properties of thin accretion discs around such BHs . Finally, we end the paper with conclusions.

\section{Geometry of galactic BHs with dark halo}
\label{2-BH}
According to the observational evidence supermassive BHs reside in the center of galaxies that dark matter surrounds them \cite{new}. In this section, we introduce the background geometry of our analysis and focus on two galactic BHs immersed in a dark halo.

\subsection{Anisotropic dark matter fluid with Hernquist-type density}
Astronomical observations and large-scale simulations describe a good and interesting galactic profile in bulges and elliptical galaxies that well defined by the Hernquist-type distribution \cite{Hernquist}. In \cite{Cardoso} authors generalized Einstein clusters using the Hernquist-type density distribution as
\begin{equation}
\rho(r)  =\frac{M a_{0}}{2\pi r (r+a_{0})^3},
\label{1}
\end{equation}
where $M$ is the mass of halo and $a_0$ is a typical lengthscale of the galaxy with the above matter distribution. The solution of the Einstein equations coupled to anisotropic dark matter fluid was obtained as follows \cite{Cardoso}
\begin{eqnarray}
ds^2&=&-\left(1-\frac{2 M_{\rm BH}}{r}\right) e^{\gamma(r)}dt^2\nonumber\\
&+&\frac{dr^2}{\left[1-\frac{2M_{\rm BH}}{r}-\frac{2M r}{(a_0+r)^2} \left(1-\frac{2 M_{\rm BH}}{r}\right)^2\right]}\nonumber\\
&+&r^2\left(d\theta^2+\sin ^2\theta d\phi^2\right),
\label{2}
\end{eqnarray}
with
\begin{equation}
\gamma(r) = -\pi \sqrt{\frac{M}{\xi}}+2 {\sqrt{\frac{M}{\xi}}}\arctan\left(\frac{r+a_0-M}{\sqrt{M \xi}}\right),
\label{5}
\end{equation}
\begin{equation}
\xi = 2a_0-M+4 M_{\rm BH}.
\label{6}
\end{equation}
Here, the density profile and tangential pressure for massive BHs at the center of Hernquist-type distribution respectively are given by
\begin{equation}
\rho(r)  =\frac{2M (a_{0}+2M_{\rm BH})(1-\frac{2M_{\rm BH}}{r})} {4\pi r (r+a_{0})^3},
\label{7}
\end{equation}

\begin{equation}
P_{t}(r)=\frac{\rho(r)}{2}\frac{\left(M_{\rm BH}+\frac{M r^2}{(a_0+r)^2} \left(1-\frac{2 M_{\rm BH}}{r}\right)^2\right)}{r-2\left(M_{\rm BH}+\frac{M r^2}{(a_0+r)^2} \left(1-\frac{2 M_{\rm BH}}{r}\right)^2\right)}.
\label{770}
\end{equation}

The above geometry describes a BH with mass $M_{\rm BH}$ immersed in a dark halo of mass $M$ and so the ADM mass $M_{\rm ADM}$ of the space-time is given by $M_{\rm ADM}=M_{\rm BH}+M$. We note the event horizon radius coincides with the Schwarzschild radius  i.e. $r_{\rm h}=2 M_{\rm BH}$. Also, by assuming a hierarchy of scales that is determined by the regime
\begin{equation}
M_{\rm BH} \ll M \ll a_0,
\label{8}
\end{equation}
the above space-time has only a curvature singularity at $r=0$. There are also curvature singularities at $r_{\pm}=M-a_{0}\pm\sqrt{M^2-2Ma_{0}-4MM_{\rm BH}}$, but such singularity does not describe an astrophysical setup since $M>2(a_{0}+2M_{\rm BH})$. Indeed, as the author has discussed in Ref.\cite{Myung:2024tkz}, when the condition (\ref{8}) is not satisfied there are two real roots $r_{\pm}$  in addition to $r_{\rm h}$; and there are two unstable photon orbits. These cases may be theoretically of interest, but they do not describe a realistic BH-halo system. Therefore, we shall consider the cases that satisfy condition (\ref{8}).
Moreover,we define the compactness parameter as
\begin{equation}
{\cal C} = \frac{M}{a_0},
\label{9}
\end{equation}
which help to compare different halo configurations and is bounded by the galactic dynamic as ${\cal C}\lesssim 10^{-4}$ \cite{C1}--\cite{C2}. Clearly, in the absence of dark halo, $M=0$, geometry (\ref{2}) describes the usual Schwarzschild space-time in GR.

\subsection{Anisotropic dark matter fluid with equation of state $P_{t}=\omega\rho$}
Similar to the method introduced by Cardoso et al. \cite {Cardoso}, a novel class of spherically symmetric and asymptotically flat BHs immersed in anisotropic dark matter fluid $T^{\mu}_{\nu} =\rm diag (\rho, 0, P_t, P_t)$ with equation of state $P_{t}(r)=\omega\rho (r)$
has been obtained in Ref. \cite {Jusufi:2022jxu}

\begin{equation}
ds^2=
\begin{cases}
-\left(1-\frac{2M_{\rm BH}}{r}\right)\left(\frac{r+8\omega M_{\rm BH}}{R+8\omega M_{\rm BH}}\right)^{4\omega}dt^2 &\\
\quad +\frac{dr^2}{\left(1-\frac{2M_{\rm BH}}{r}\right)\left[1-\frac{4\omega}{1+4\omega}\left(1-\frac{2M_{\rm BH}}{r}\right)\right]}+r^2\left(d\theta^2+\sin ^2\theta d\phi^2\right), &r<R \\\\
-\left(1-\frac{2M}{r}\right)dt^2+\frac{dr^2}{\left(1-\frac{2M}{r}\right)}&r\geq R\\
\quad +r^2\left(d\theta^2+\sin ^2\theta d\phi^2\right), &
\end{cases}\label{n1}
\end{equation}

where
\begin{equation}
M = M_{\rm BH}+\frac{2\omega R}{1+4\omega}\left(1-\frac{2M_{\rm BH}}{R}\right)^2,
\label{n2}
\end{equation}
in which the interior region of the galaxy is described by the line element in region $r<R$, and the exterior region $r\geq R$ is describe by the Schwarzschild space-time geometry.

Note that in the case of $\omega=0$, the above line-element reduces to the spherically symmetric vacuum solutions in the absence of dark matter fluid. Moreover, similar to the Cardoso’s BH, this solution has a regular event horizon at $r_{\rm h} = 2 M_{\rm BH}$. In what follows we shall discuss the geodesic motion of gas/accretion particles around such BHs in the interior region of galaxy $r<R$.

From the EHT results for the shadow radius of Sgr A* within $2\sigma$ confidence level, one can constrain the state parameter $\omega$ as $\omega \leq 10^{-3}$. The $\omega$ parameter used here is consistent with shadow image of the Sgr A*. We shall also set $R=10^{16}M_{\rm BH}$ and $M_{\rm BH}=4.1\times 10^6M_{\odot}$ \cite{Jusufi:2022jxu}.

\section{Geodesic equations}
\label{3-geodesic}
When diffuse material moving along geodesic paths around a central compact object an accretion disk is formed. In this section we study the trajectory of test particles in the equatorial plane of a thin accretion disk which is governed by the Lagrangian
\begin{equation}
{\cal L}= \frac{1}{2}g_{\mu\nu}\dot{x^{\mu}}\dot{x^{\nu}},
\label{g1}
\end{equation}
where $g_{\mu\nu}$ is the space-time metric and a dot denotes differentiation with respect to the affine parameter. The metric of a general static and spherically symmetric space-time is expressed as
\begin{equation}
ds^2=g_{tt}dt^2+g_{rr}dr^2+g_{\theta\theta}d\theta ^2+g_{\phi\phi}d\phi ^2,
\label{g2}
\end{equation}
where $g_{tt}, g_{rr}, g_{\theta\theta}$ and $g_{\phi\phi}$ components only depend on the radial coordinate $r$. Now using the Euler-Lagrange equations, in the equatorial plane $\theta=\frac{\pi}{2}$, the geodesic equations for $t$ and $\phi$ coordinates can be obtained as follows
\begin{equation}
p_{t}=\dot{t}=-\frac{\tilde{E}}{g_{tt}},
\label{g3}
\end{equation}
\begin{equation}
p_{\phi}=\dot{\phi}=\frac{\tilde{L}}{g_{\phi\phi}},
\label{g4}
\end{equation}
where $\tilde{E}$ and $\tilde{L}$ are the particle's specific energy and the specific angular momentum, respectively. Then, by substituting (\ref{g3}) and (\ref{g4}) in Eq. (\ref{g1}), and using condition $2{\cal L}=-1$ for test particles we find
\begin{equation}
-g_{tt}g_{rr}\dot{r}^2+V_{\rm eff}(r)=\tilde{E}^2,
\label{g5}
\end{equation}
where the effective potential is given by
\begin{equation}
V_{\rm eff}(r)=-g_{tt}\left(1+\frac{\tilde{L}^2}{g_{\phi\phi}}\right).
\label{g6}
\end{equation}

The stable circular orbits in the equatorial plane obey the conditions $V_{\rm eff}(r)=0$ and $V_{\rm eff,r}(r)=0$. Using these conditions, one can find the specific energy, the specific angular momentum as well as the angular velocity, $\Omega=d\phi/dt$, of particles moving in the gravitational potential of the compact object as follows
\begin{equation}
{\tilde{E}}=-\frac{g_{tt}}{\sqrt{-g_{tt}-g_{\phi\phi}\Omega^2}},
\label{g7}
\end{equation}
\begin{equation}
{\tilde{L}}=\frac{g_{\phi\phi}\Omega}{\sqrt{-g_{tt}-g_{\phi\phi}\Omega^2}},
\label{g8}
\end{equation}
\begin{equation}
\Omega=\sqrt{\frac{-g_{tt,r}}{g_{\phi\phi,r}}}.
\label{g9}
\end{equation}
Furthermore, to obtain the radius of the innermost stable circular orbit (ISCO) we use the condition $V_{\rm eff,rr}(r)=0$ which leads to the following equation
\begin{equation}
{\tilde{E}^2g_{\phi\phi,rr}}+{\tilde{L}^2g_{tt,rr}}+(g_{tt}g_{\phi\phi})_{,rr}=0.
\label{g10}
\end{equation}

Since the above quantities depend only on the space-time metric components, in the next section we obtain them for galactic BHs immersed in dark matter halo (\ref{2}) and (\ref{n1}).

\section{Observational properties of thin accretion disks}
\label{4-disk}
Now, we consider the steady-state Novikov-Thorne model to study the accretion process in thin disks around central galactic BHs immersed in a dark matter halo and explore the effects of dark halo on the observational properties of disks in such a space-time. The model is based on several assumptions: the thin disk is identified by the fact that its vertical size, $h$, is much smaller than its horizontal size, $h\ll r$. The thin disk is situated at the equatorial plane and its particles move in Keplerian orbits between the inner edge on the ISCO radius and the outer edge $r_{\rm out}$. The space-time is assumed to be stationary, axisymmetric, asymptotically flat and the impact of the disk’s mass on the background metric is ignored. The model also consider the disk is in a steady-state meaning that the accretion rate, $\dot{M}_{0}$, is constant over time. Moreover, the accreting matter is assumed to be in thermal equilibrium.


In the steady-state accretion disk model the energy-momentum tensor of accreting matter can be written as \cite{Novikov}
\begin{equation}
T^{\mu\nu}=\rho_0u^{\mu}u^{\nu}+2u^{(\mu}q^{\nu)}+t^{\mu\nu},
\label{g11}
\end{equation}
where $u_{\mu}q^{\mu}=0$ and $u_{\mu}t^{\mu\nu}=0$. $u^{\mu}$ is the four-velocity of orbiting particles, while the rest mass density, the energy flow vector and stress tensor of accreting matter are denoted by $\rho_0$, $q^{\mu}$ and $t^{\mu\nu}$, respectively. From the conservation laws of mass $\nabla_{\mu}(\rho_0u^{\mu})=0$, energy $\nabla_{\mu}E^{\mu}=0$, and angular momentum $\nabla_{\mu}J^{\mu}=0$, we find the time-averaged radial structure equations of disk as follows
\begin{equation}
\dot{M}_{0}=-2\pi\sqrt{-g}\Sigma u^r=\rm const,
\label{g12}
\end{equation}
\begin{equation}
[\dot{M}_{0}\tilde{E}-2\pi\sqrt{-g}\Omega W^{r}_{\phi}]_{,r}=4\pi rF(r)\tilde{E},
\label{g13}
\end{equation}
\begin{equation}
[\dot{M}_{0}\tilde{L}-2\pi\sqrt{-g}W^{r}_{\phi}]_{,r}=4\pi rF(r)\tilde{L},
\label{g14}
\end{equation}
where $\sqrt{-g}$ is the metric determinant, and the time averaged surface density $\Sigma(r)$, and the averaged torque $W^{r}_{\phi}$ are given by
\begin{equation}
\Sigma(r)=\int_{-h}^{h}\langle\rho_0\rangle dz,
\label{g15}
\end{equation}
\begin{equation}
W^{r}_{\phi}=\int_{-h}^{h}\langle t^{r}_{\phi}\rangle dz.
\label{g16}
\end{equation}
where $z$ is the cylindrical coordinate and $\langle t^{r}_{\phi}\rangle$ is the average value of $(\phi,r)$ component of the stress tensor on the time scale $\Delta t$ and azimuthal angle $\Delta\phi=2\pi$. Then, using the energy-angular momentum relation as $\tilde{E}_{,r}=\Omega\tilde{L}_{,r}$ and removing $W^{r}_{\phi}$ from Eqs. (\ref{g13}) and (\ref{g14}), the time averaged energy flux emitted from the disk surface is given by \cite{Novikov}
\begin{equation}
F(r)=-\frac{\dot{M}_{0}\Omega_{,r}}{4\pi\sqrt{-g}\left(\tilde{E}-\Omega \tilde{L}\right)^2}\int^r_{r_{\rm isco}}\left(\tilde{E}-\Omega \tilde{L}\right) \tilde{L}_{,r}dr.
\label{g17}
\end{equation}

Due to the assumption of thermodynamic equilibrium in the steady-state thin disk model, the emitted radiation from the disk surface can be described as a perfect black body radiation. Therefore, the temperature of disk is related to the energy flux through the following equation
\begin{equation}
F(r)=\sigma_{\rm SB} T^4(r),
\label{g18}
\end{equation}
where $\sigma_{\rm SB}=5.67\times10^{-5}\rm erg$ $\rm s^{-1} cm^{-2} K^{-4}$ is the Stefan-Boltzmann constant. Moreover, the observed luminosity has also a redshifted
black body spectrum as follows \cite{Torres}
 \begin{equation}
L(\nu)=\frac{8\pi h \cos\gamma}{c^2}\int_{r_{\rm in}}^{r_{\rm out}}\int_0^{2\pi}\frac{ \nu_{\rm e}^3 r dr d\phi}{\exp{[\frac{h\nu_{\rm e}}{k_{\rm B} T}]}-1},
\label{g19}
\end{equation}
where $h$ is the Planck constant, and $k_{\rm B}$ is the Boltzmann constant. $\gamma$ represents the inclination angle of accretion disk, and $r_{\rm in}$ and $r_{\rm out}$ are the inner and outer radii of the edge of disk, respectively. Also, the emitted frequency is defined as $\nu_{\rm e}=\nu (1+z)$ and the redshift factor is given by
\begin{equation}
1+z=\frac{1+\Omega r\sin\phi\sin\gamma}{\sqrt{-g_{tt}-\Omega ^2g_{\phi\phi}}}.
\label{g20}
\end{equation}

\begin{figure*}[htbp]
  \centering
  \includegraphics[width=7.5cm,height=5cm]{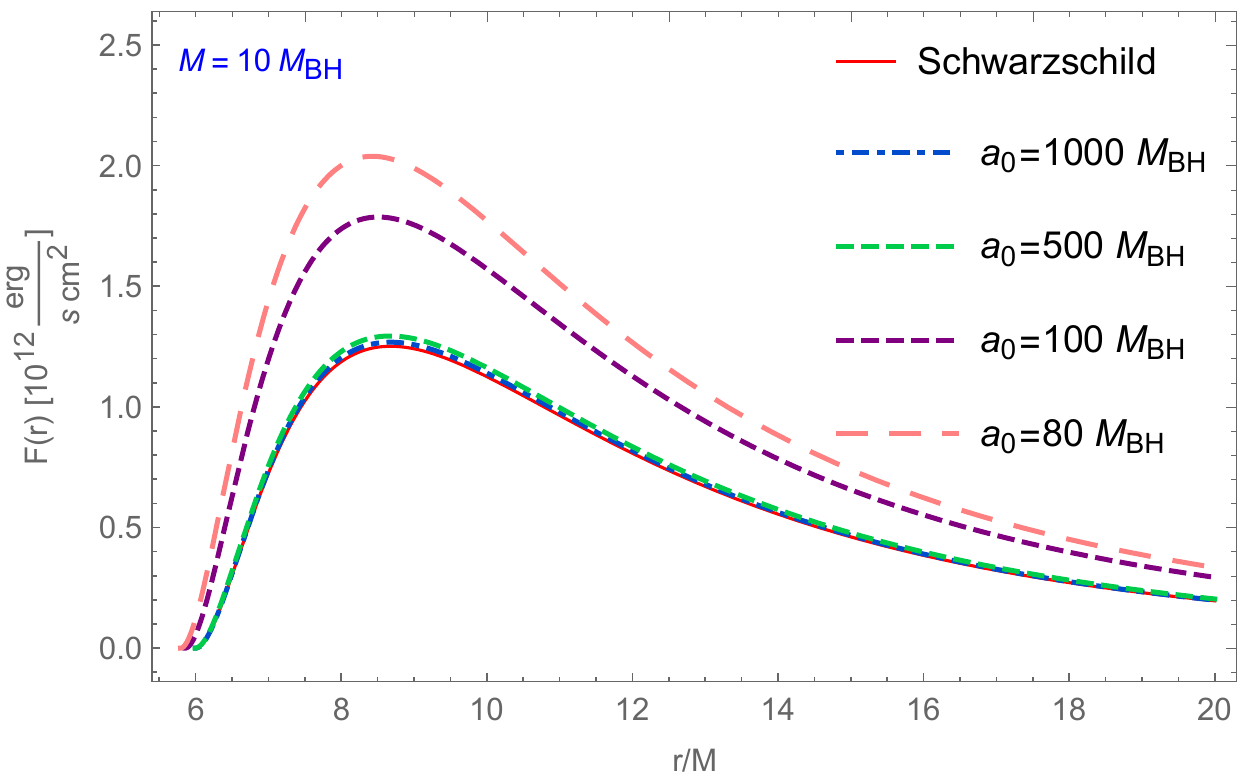}
  \hspace{0.1cm}
  \includegraphics[width=7.5cm,height=5cm]{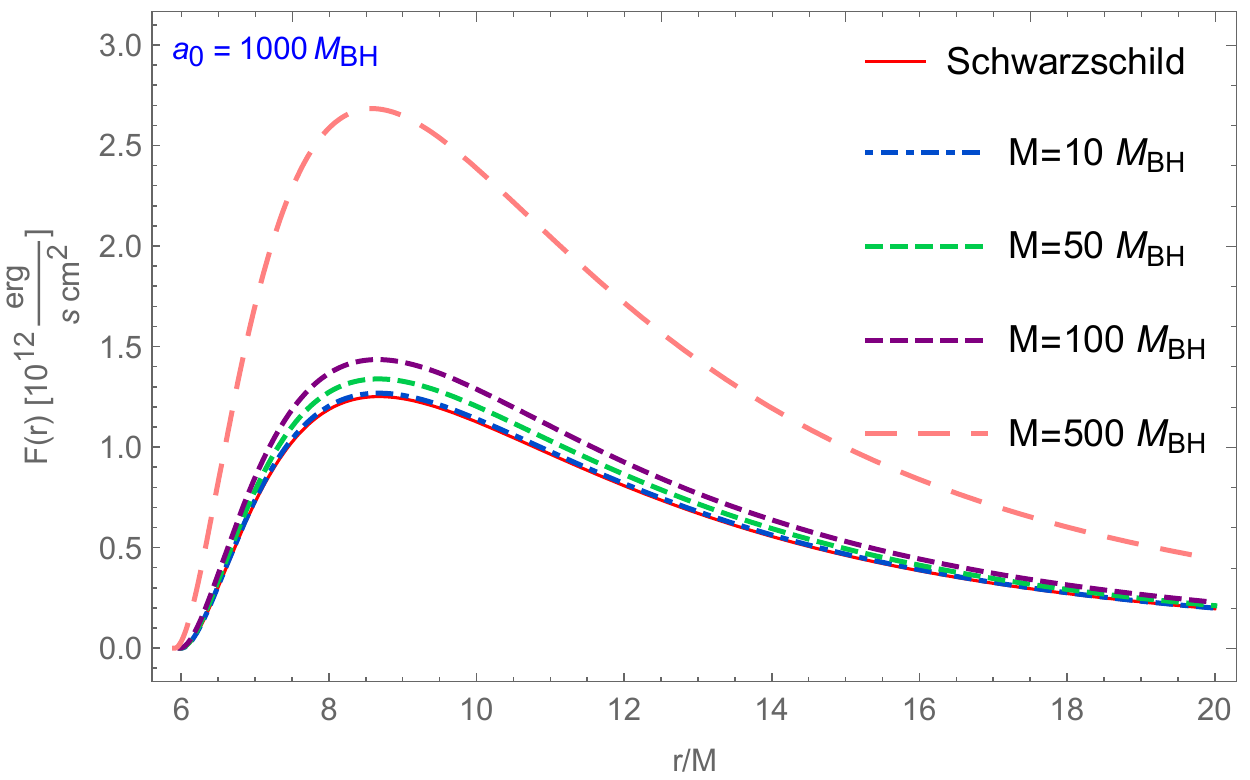}
  \caption {\footnotesize The energy flux $F(r)$ of an accretion disk around a static BH at the center of Hernquist-type density distribution with the mass accretion rate $\dot{M}=2\times10^{-6}M_{\odot}yr^{-1}$, for different values of $a_0$, left panel and different values of $M$, right panel. In each panel the solid red curve correspond to the Schwarzschild BH without dark matter.}\label{f}
\end{figure*}

\begin{figure*}[htbp]
  \centering
  \includegraphics[width=7.5cm,height=5cm]{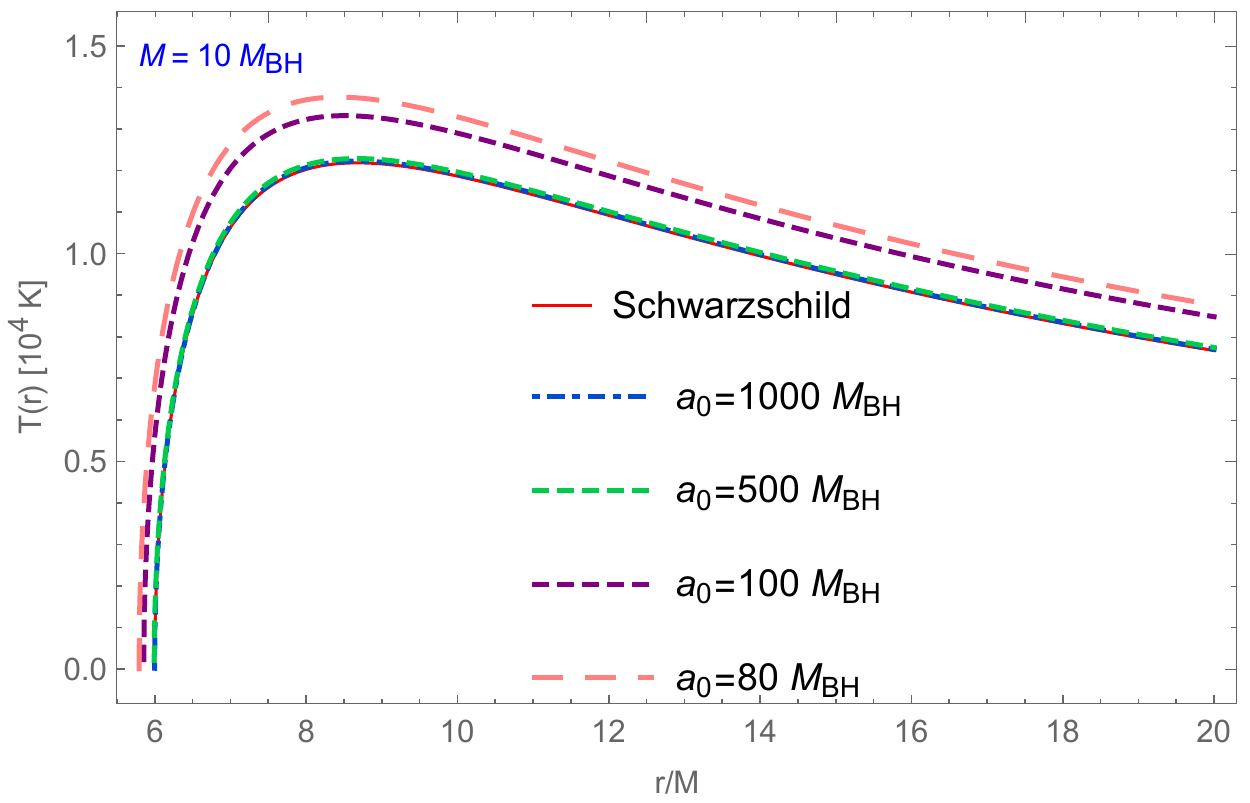}
  \hspace{0.1cm}
  \includegraphics[width=7.5cm,height=5cm]{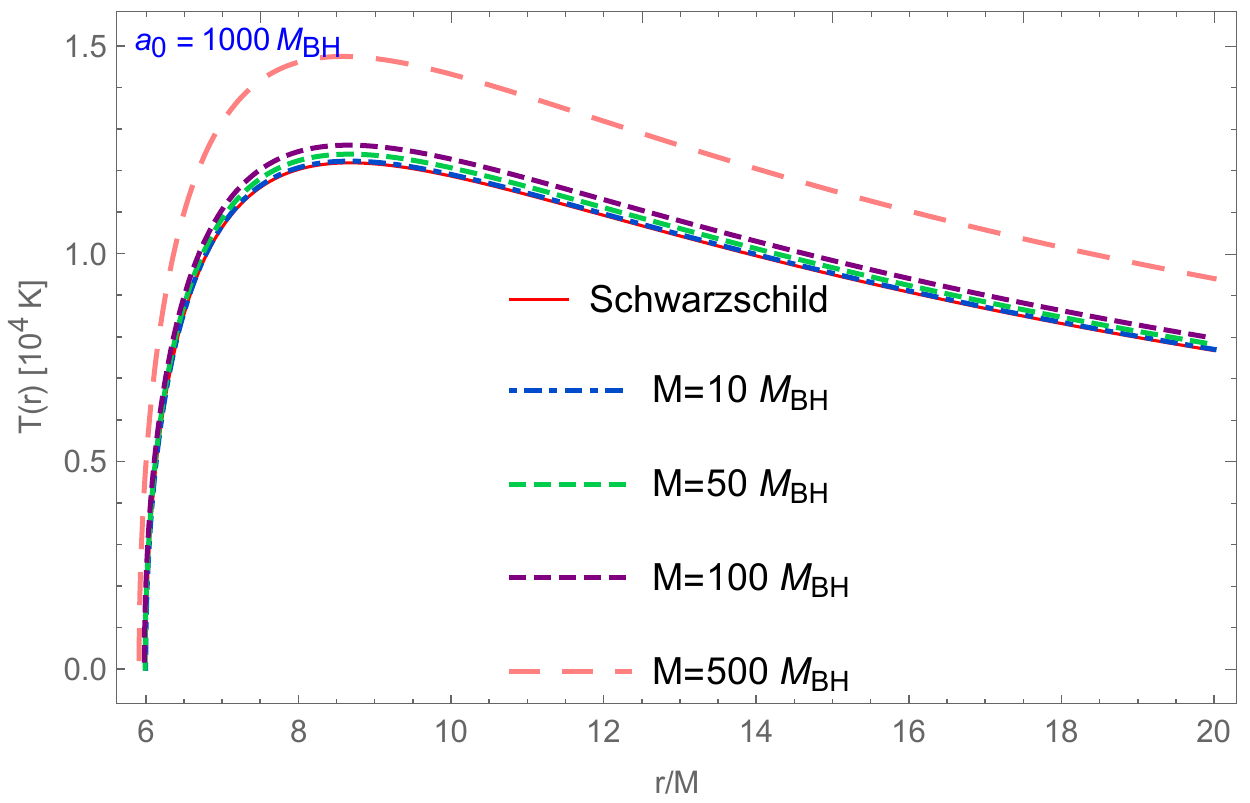}
  \caption {\footnotesize \footnotesize Disk temperature $T(r)$ for a static BH at the center of Hernquist-type density distribution with mass accretion rate $\dot{M}=2\times10^{-6}M_{\odot}yr^{-1}$, for different values of $a_0$, left panel and different values of $M$, right panel. In each panel the solid red curve correspond to the Schwarzschild BH without dark matter.}\label{t}
\end{figure*}

\begin{figure*}[htbp]
  \centering
  \includegraphics[width=7.5cm,height=5cm]{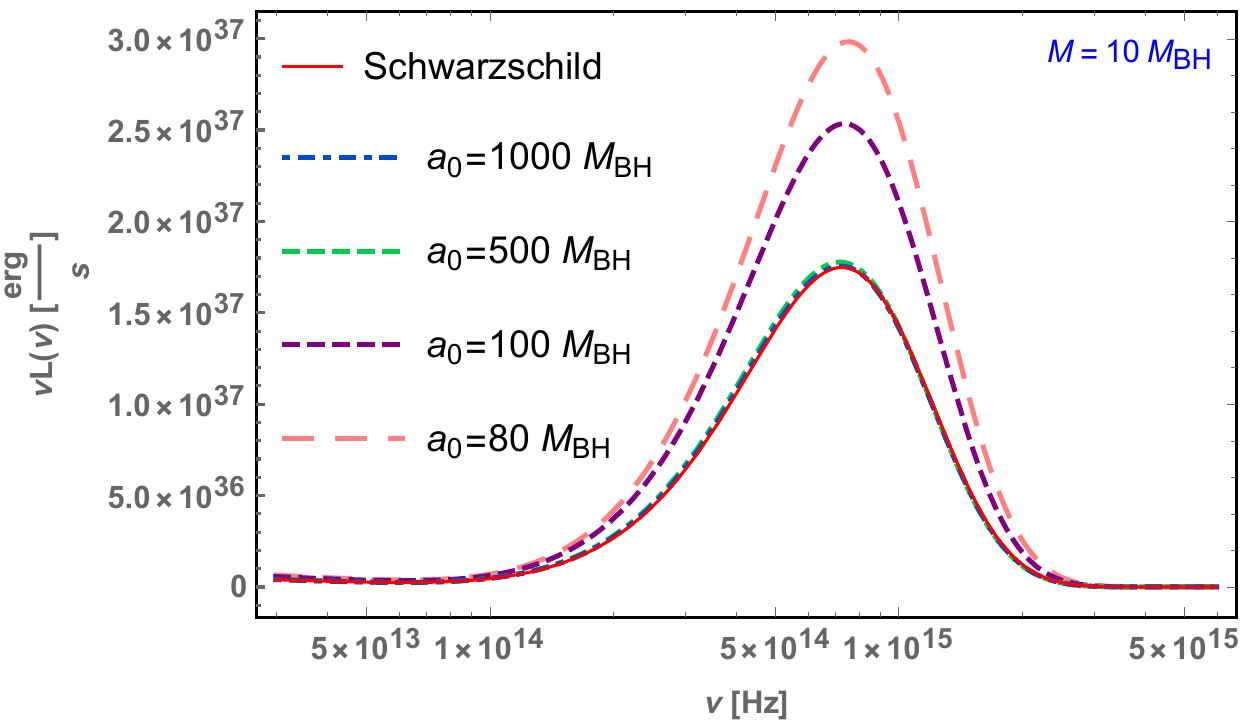}
  \hspace{0.1cm}
  \includegraphics[width=7.5cm,height=5cm]{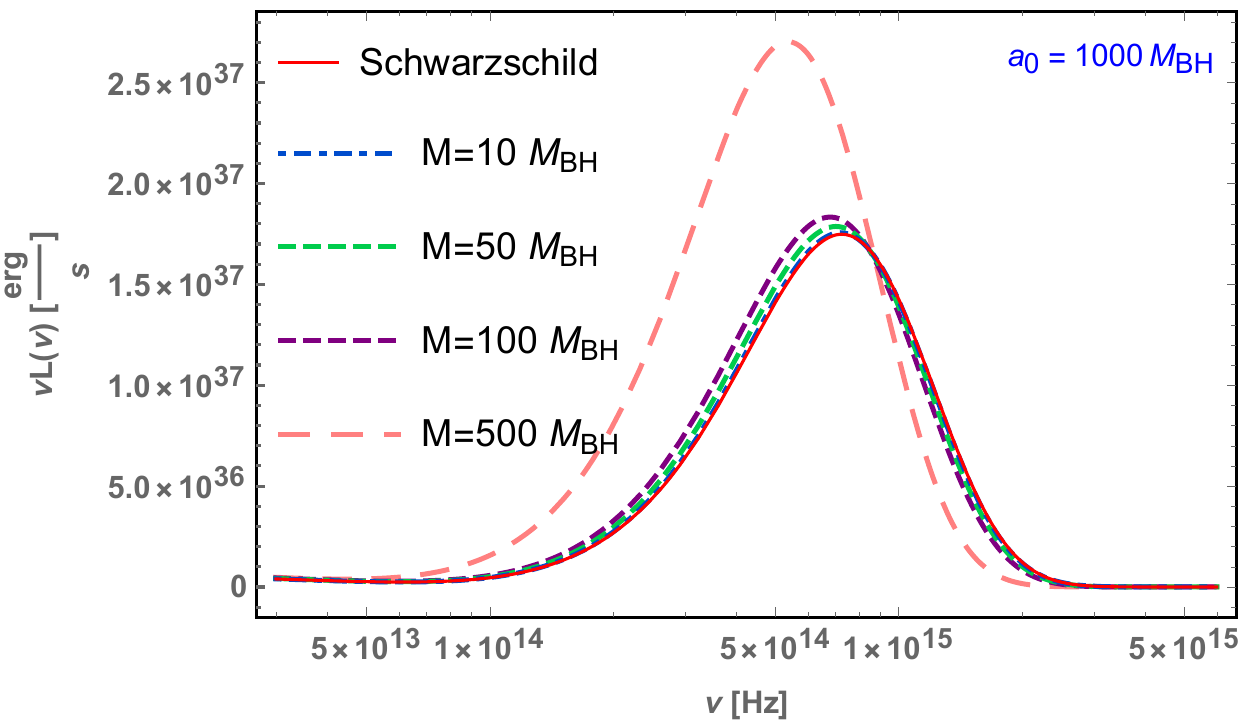}
  \caption {\footnotesize The emission spectrum $\nu L(\nu)$ of an accretion disk around a static BH at the center of Hernquist-type density distribution with mass accretion rate $\dot{M}=2\times10^{-6}M_{\odot}yr^{-1}$ and inclination $\gamma=0^{\circ}$ as a function of frequency $\nu$ for different values of $a_0$, left panel and different values of $M$, right panel. In each panel the solid red curve correspond to the Schwarzschild BH without dark matter.}\label{l}
\end{figure*}

We numerically calculate the radiant energy flux $F(r)$, the disk temperature $T(r)$ and the emission spectra $L(\nu)$ for accretion disk around galactic BHs immersed in dark halo and plot them in the left panel of Figs. ~\ref{f}, ~\ref{t} and ~\ref{l}, respectively for a fixed value of $M$ and different $a_{0}$. We also show a comparison with the corresponding results for the Schwarzschild BH without dark matter. It can be seen from the figures that by turning on the effects of dark matter and increasing the compactness parameter, $M/a_0$, the energy flux, temperature and luminosity increase too. So, the disk around Schwarzschild BH is cooler and less efficient than the BHs immersed in dark matter halo. Moreover, we find that the effect of dark halo on the disk properties is larger if the compactness of dark matter halo is bigger. Similar results can be seen from right panel of figures for a given value of $a_0$ and different $M$.

The accretion efficiency is another important quantity in the mass accretion process which describes the capability of central object in transmuting rest mass into outgoing radiation. In the Novikov-Thorne model by assuming that all emitted photons can escape to infinity, one can obtain the radiative efficiency $\eta$ using the specific energy of  particles at ISCO radius \cite{Novikov}
\begin{equation}
\eta=1-\tilde{E}_{\rm isco}.
\label{g21}
\end{equation}

In Table ~\ref{TT1} and ~\ref{TT2}, we present the numerical values of $r_{\rm isco}$, the maximum values of energy flux, the disk temperature, emission spectra and also the radiative efficiency of BHs in dark halo for different values of $M$ and $a$ parameters.

\begin{table*}
\centering
\caption{\footnotesize The numerical results for $r_{\rm{isco}}$, the maximum of energy flux, maximum of temperature, the critical frequency, maximum emission spectrum, and efficiency of an accretion disk around a static BH at the center of Hernquist-type density distribution, with $M=10$ and different values of $a_0$. The last row corresponds to the Schwarzschild BH without dark matter.}
\begin{tabular}{|l|l|l|l|l|l|l|l|l|}
\hline
$\shortstack{$M/M_{\rm BH}$ \\ $$}$ &$\shortstack{$a_{0}/M_{\rm BH}$ \\ $$}$& $\shortstack{$r_{\rm isco}/M_{\rm BH}$ \\ $$}$& \shortstack{$F_{\rm max}(r)$ \\ $[\rm erg$ $\rm s^{-1} cm^{-2}]\times10^{12}$}& \shortstack{$T_{\rm max}(r)$ \\ $[\rm K]\times10^{4}$}& \shortstack{$\nu_{\rm c}$ \\ $[\rm Hz]\times10^{14}$}& \shortstack{$\nu L(\nu)_{\rm max}$ \\ $[\rm erg]\times10^{37}$}& $\shortstack{$\eta$ \\ $\quad\quad\quad$}$\\ [0.5ex]
\hline\hline
{\quad10}  &\quad80& \quad5.7930&\quad\quad\quad2.0386&\quad1.3770&\quad7.5341&\quad2.9836& \quad0.1606\\

{\quad10}
     &\quad100& \quad5.8552&\quad\quad\quad1.7867&\quad1.3323 &\quad7.3691&\quad2.5348&\quad0.1419\\

{\quad10}
    &\quad500& \quad5.9926&\quad\quad\quad1.2937&\quad1.2290&\quad7.2895&\quad1.7774&\quad0.0756\\

{\quad10}
    &\quad1000&\quad 5.9981&\quad\quad\quad1.2680&\quad1.2229&\quad7.2188&\quad1.7555&\quad0.0665\\

{\quad-}
&\quad-&\quad 6&\quad\quad\quad1.2511&\quad1.2188&\quad7.1908&\quad1.7483&\quad0.0572\\
\hline
\end{tabular}
\label{TT1}
\end{table*}

\begin{table*}
\centering
\caption{\footnotesize The numerical results for $r_{\rm{isco}}$, the maximum of energy flux, maximum of temperature, the critical frequency, maximum emission spectrum, and efficiency of an accretion disk around a static BH at the center of Hernquist-type density distribution, with $a_0=1000$ and different values of $M$. The last row corresponds to the Schwarzschild BH without dark matter.}
\begin{tabular}{|l|l|l|l|l|l|l|l|l|}
\hline
$\shortstack{$M/M_{\rm BH}$ \\ $$}$&$\shortstack{$a_{0}/M_{\rm BH}$ \\ $$}$& $\shortstack{$r_{\rm isco}/M_{\rm BH}$ \\ $$}$& \shortstack{$F_{\rm max}(r)$ \\ $[\rm erg$ $\rm s^{-1} cm^{-2}]\times10^{12}$}& \shortstack{$T_{\rm max}(r)$ \\ $[\rm K]\times10^{4}$}& \shortstack{$\nu_{\rm c}$ \\ $[\rm Hz]\times10^{14}$}& \shortstack{$\nu L(\nu)_{\rm max}$ \\ $[\rm erg]\times10^{37}$}& $\shortstack{$\eta$ \\ $\quad\quad\quad$}$\\ [0.5ex]
\hline\hline
{\quad500}  &\quad1000&\quad 5.9118&\quad\quad\quad2.6835&\quad1.4749&\quad7.5316&\quad2.7038& \quad0.4821\\

{\quad100}
     &\quad1000& \quad5.9814&\quad\quad\quad1.4351&\quad1.2613&\quad7.3824&\quad1.8336&\quad0.1490\\

{\quad50}
    &\quad1000&\quad 5.9906&\quad\quad\quad1.3389&\quad1.2396&\quad7.2360&\quad1.7869&\quad0.1035\\

{\quad10}
    &\quad1000&\quad 5.9981&\quad\quad\quad1.2680&\quad1.2229&\quad7.2188&\quad1.7555&\quad0.0665\\

{\quad-}
&\quad-& \quad6&\quad\quad\quad1.2511&\quad1.2188&\quad7.1908&\quad1.7483&\quad0.0572\\
\hline
\end{tabular}
\label{TT2}
\end{table*}

\begin{figure*}[htbp]
  \centering
  \includegraphics[width=7.5cm,height=5cm]{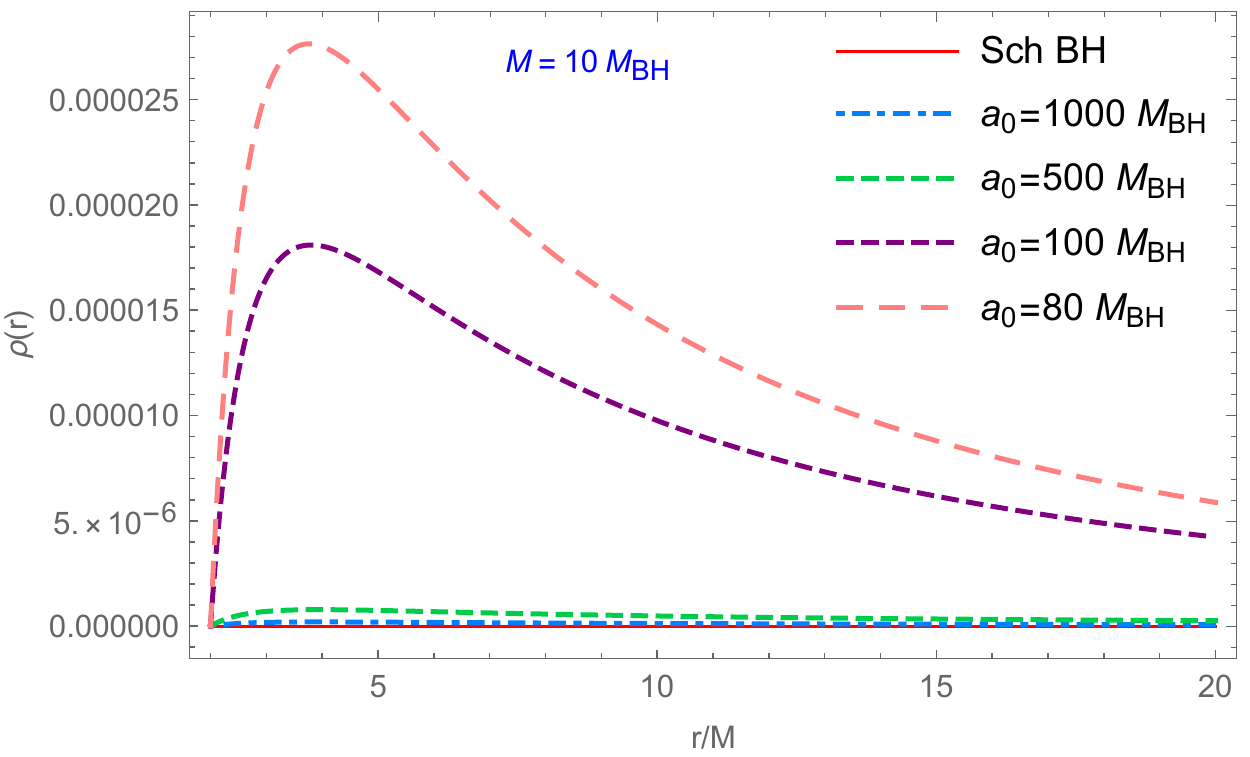}
  \hspace{0.1cm}
  \includegraphics[width=7.5cm,height=5cm]{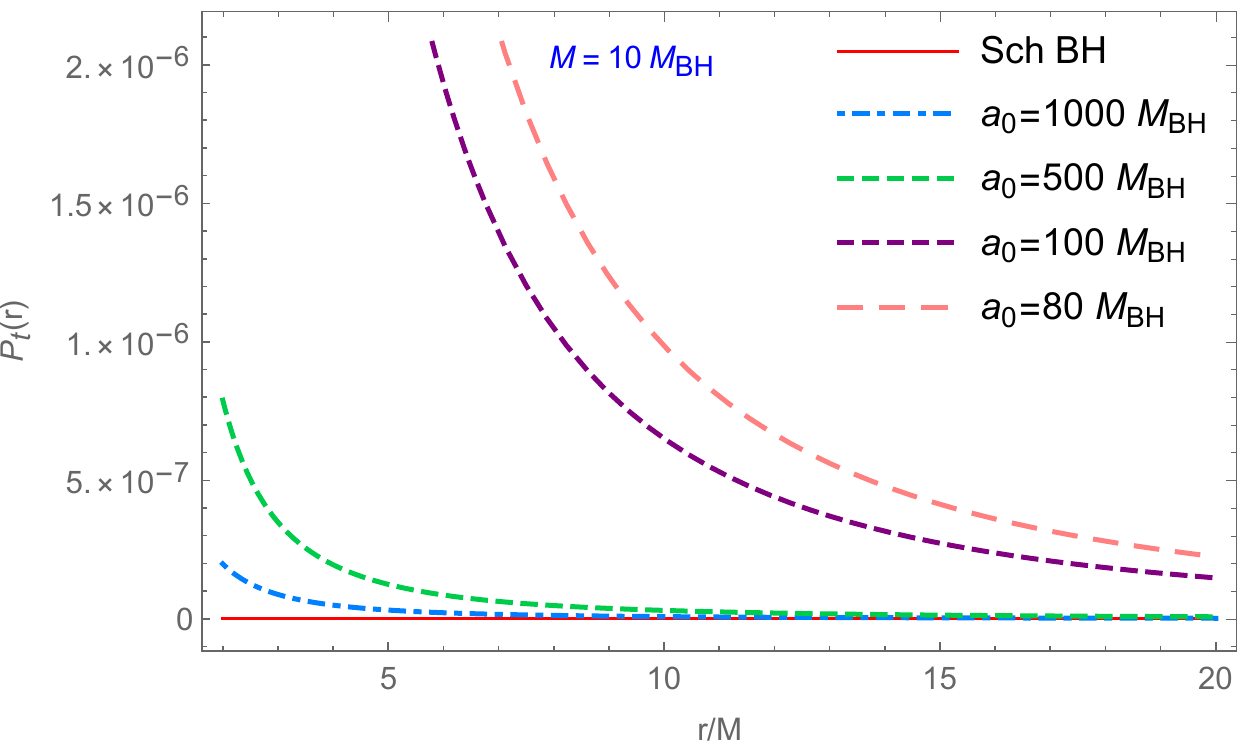}
  \caption {\footnotesize The density profile (left panel) and tangential pressure (right panel) of BHs at the center of Hernquist-type density distribution as a function of $r$ for $M=10$ and different values of $a_0$. In each panel the solid red curve correspond to the Schwarzschild BH without dark matter.}\label{rp}
\end{figure*}

The Cardoso BHs we consider here, have the density profile and the tangential pressure as given in Eqs. ~(\ref{7}) and (\ref{770}). The dependence of the matter density and tangential pressure on the parameter $a_{0}$ is shown in the left and right panels of Fig. ~\ref{rp}, respectively. As is clear from the figure, the BH with a larger compactness parameter, ${\cal C}=M/a_{0}$, has a larger density and tangential pressure. So it is expected that by increasing the tangential pressure which mimics the dark matter’s angular velocity, the position of ISCO radius is shifted towards the smaller radii which is in agreement with Table ~\ref{TT1}. The decrease of ISCO radius leads to the decrease of disk surface and thus increase the radiation flux in accordance with Fig. ~\ref{f}.

\begin{figure*}[htbp]
  \centering
  \includegraphics[width=7.5cm,height=5cm]{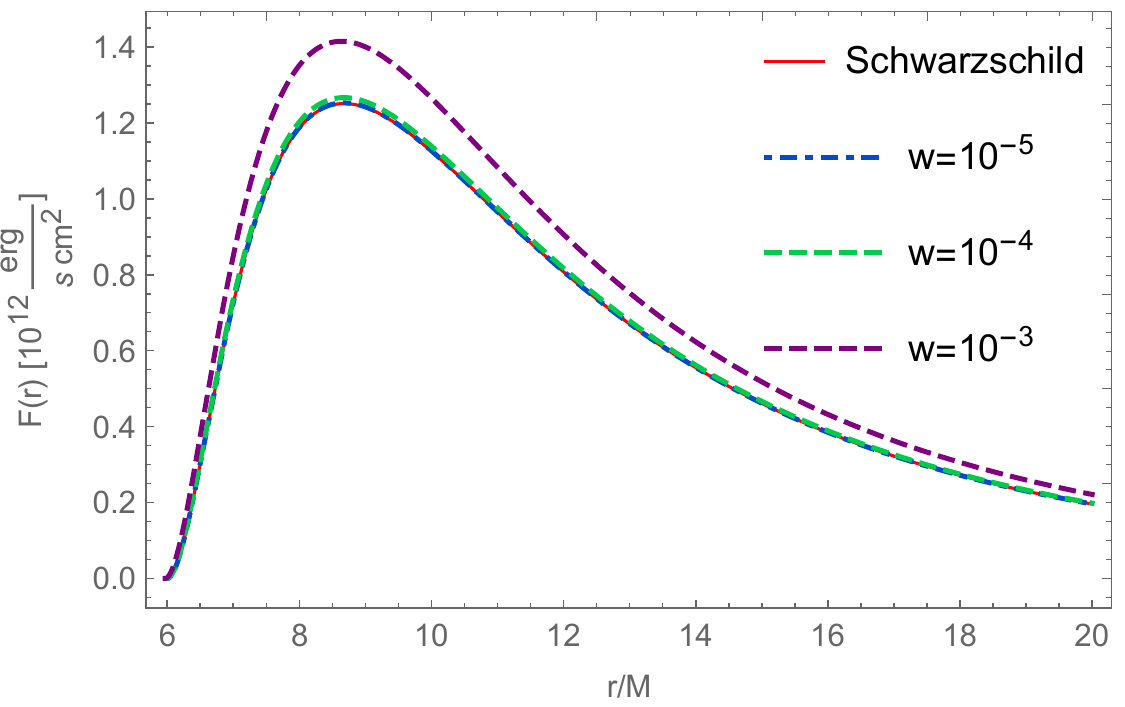}
  \hspace{0.1cm}
  \includegraphics[width=7.5cm,height=5cm]{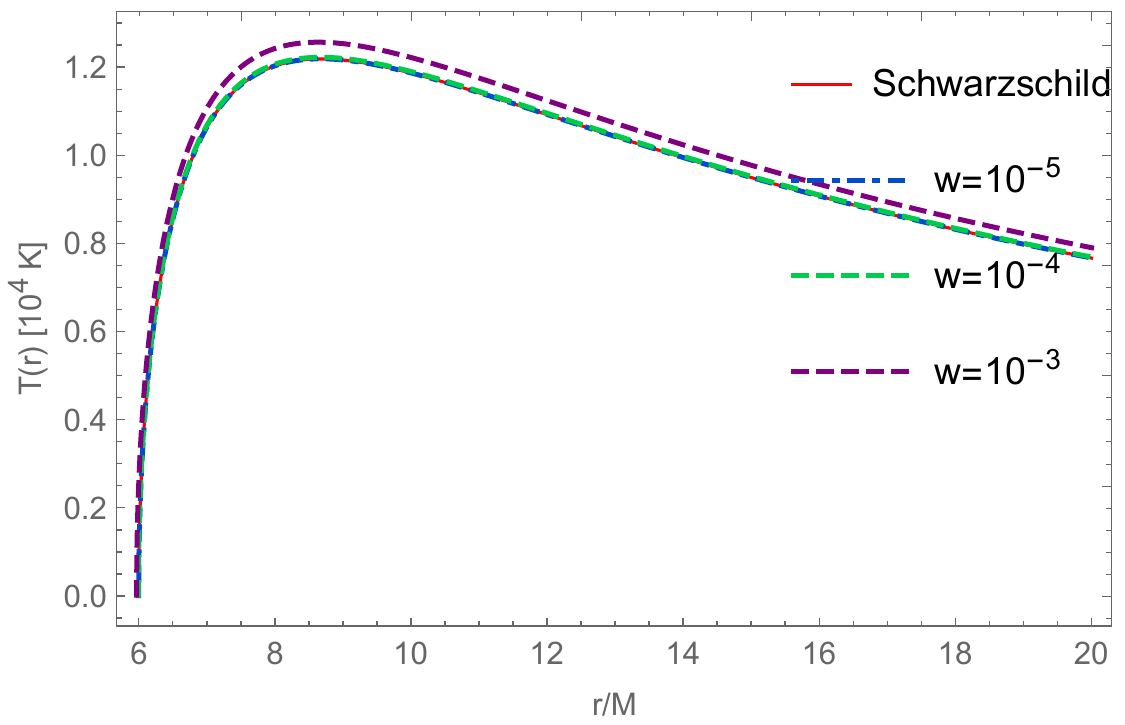}
  \caption {\footnotesize The energy flux $F(r)$ (left panel), and the temperature $T(r)$ (right panel) of an accretion disk around a static BH immersed in anisotropic dark matter fluid with the mass accretion rate $\dot{M}=2\times10^{-6}M_{\odot}yr^{-1}$, for different values of $\omega$. We set $M_{\rm BH}=4.1\times 10^6M_{\odot}$ and $R=10^{16}M_{\rm BH}$. In each panel the solid red curve correspond to the Schwarzschild BH without dark matter.}\label{Jf}
\end{figure*}

\begin{figure}[H]
\centering
\includegraphics[width=3in]{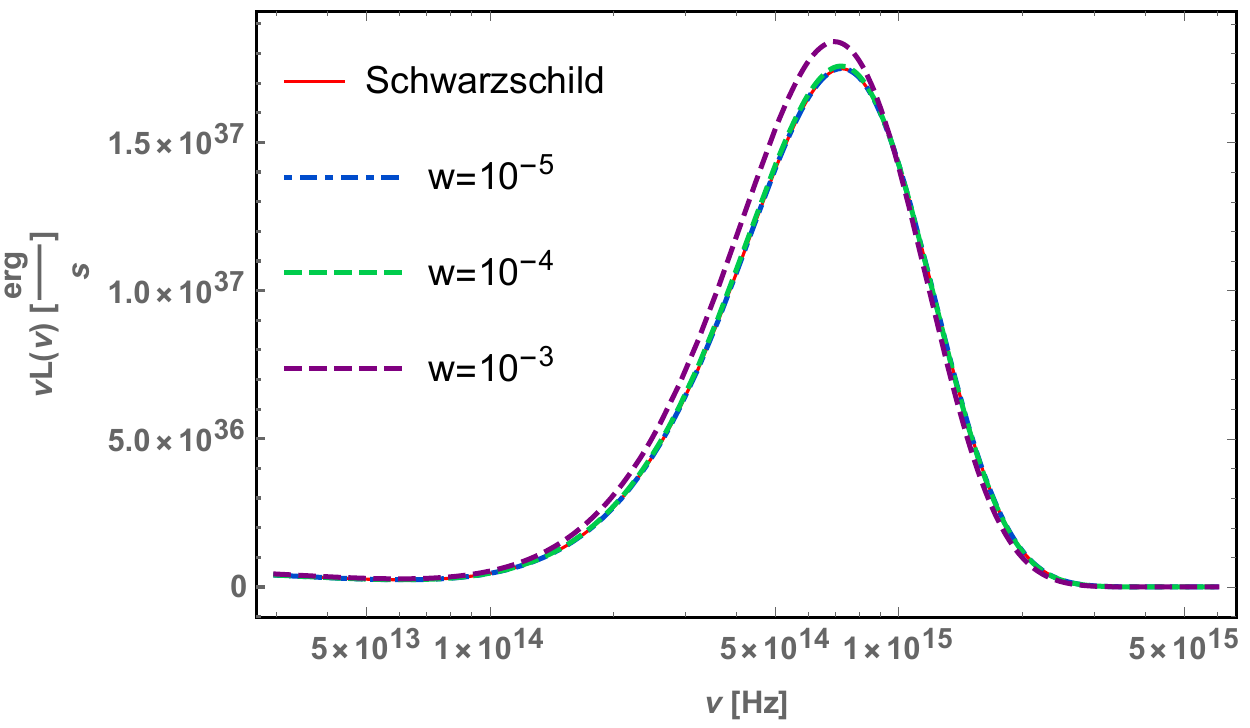}
\caption{\footnotesize The emission spectrum $\nu L(\nu)$ of an accretion disk around a static BH immersed in anisotropic dark matter fluid with mass accretion rate $\dot{M}=2\times10^{-6}M_{\odot}yr^{-1}$ and inclination $\gamma=0^{\circ}$ as a function of frequency $\nu$ for different values of $\omega$. We set $M_{\rm BH}=4.1\times 10^6M_{\odot}$ and $R=10^{16}M_{\rm BH}$. In each panel the solid red curve correspond to the Schwarzschild BH without dark matter.}
\label{Jl}
\end{figure}

Next, we consider the BH space-time with metric function given by Eq. (\ref{n1}) and obtain the electromagnetic properties of thin accretion disk around it. The dependence of the radiative flux and the disk temperature on the state parameter $\omega$ is shown in the left and right panels of Fig. ~\ref{Jf}, respectively. The figure shows that by increasing $\omega$ in the interval $\omega\leq 10^{-3}$, the radiative flux and the disk temperature increase compared to the Schwarzschild BH with $\omega=0$. However, we see that the parameter $\omega$ slightly increases the properties of disk, and thus causes that the BHs in Jusufi's model to be less distinguishable from the Schwarzschild BH than the BHs in Cardoso's model. The same behavior can be seen for the emission spectrum in Fig. ~\ref{Jl}.

Furthermore, we summarized the numerical results of $r_{\rm isco}$, the maximum values of radiative flux, the disk temperature, the luminosity and also the radiative efficiency of BHs in dark halo for different values of $\omega$ in Table ~\ref{TT3}. As can be seen, by increasing $\omega$ the ISCO radius of BHs decreases which leads to the increase of the maximum values of radiation energy and the disk luminosity in agreement with Figs. ~\ref{Jf} and ~\ref{Jl}.

\begin{table*}
\centering
\caption{\footnotesize The numerical results for $r_{\rm{isco}}$, the maximum of energy flux, maximum of temperature, the critical frequency, maximum emission spectrum, and efficiency of an accretion disk around a static BH immersed in anisotropic dark matter fluid, with $a_0=1000$ and different values of $M$. The last row corresponds to the Schwarzschild BH without dark matter.}
\begin{tabular}{|l|l|l|l|l|l|l|l|}
\hline
$\shortstack{$\omega$ \\ $\quad\quad\quad$}$& $\shortstack{$r_{\rm isco}/M_{\rm BH}$ \\ $$}$& \shortstack{$F_{\rm max}(r)$ \\ $[\rm erg$ $\rm s^{-1} cm^{-2}]\times10^{12}$}& \shortstack{$T_{\rm max}(r)$ \\ $[\rm K]\times10^{4}$}& \shortstack{$\nu_{\rm c}$ \\ $[\rm Hz]\times10^{14}$}& \shortstack{$\nu L(\nu)_{\rm max}$ \\ $[\rm erg]\times10^{37}$}& $\shortstack{$\eta$ \\ $\quad\quad\quad$}$\\ [0.5ex]
\hline\hline

$\quad10^{-3}$& \quad5.9689&\quad\quad\quad1.4151&\quad1.2569&\quad7.2315&\quad1.8396& \quad0.1198\\

$\quad10^{-4}$& \quad5.9968&\quad\quad\quad1.2667&\quad1.2226&\quad7.2102&\quad1.7575&\quad0.0636\\

$\quad10^{-5}$& \quad5.9997&\quad\quad\quad1.2526&\quad1.2192&\quad7.2034&\quad1.7493&\quad0.0578\\

\quad0& \quad6&\quad\quad\quad1.2511&\quad1.2188&\quad7.1908&\quad1.7483&\quad0.0572\\
\hline
\end{tabular}
\label{TT3}
\end{table*}

\section{Conclusions}
Recently, Cardoso et al. have generalized Einstein cluster to construct a geometry of space-time by solving Einstein’s equations minimally coupled to anisotropic dark matter fluid describing a supermassive BH immersed in a dark halo with the Hernquist-type density profile \cite {Cardoso}. In a similar way, depending on the specific choice of the mass profile, Jusufi has obtained two special solutions including an asymptotically flat BH and a naked singularity in the center of galaxy surrounded by anisotropic dark fluid with equation of state $P_{t}=\omega\rho$. \cite {Jusufi:2022jxu}.

In this paper, by considering these two types of BHs, we have estimated the value of luminosity of BH's accretion disk. For this purpose, we have studied the circular motion of test particles and derived the ISCO radius. For two BH solutions with zero radial pressure, we investigated the influence of the tangential pressure on the location of ISCO radius and thus on the electromagnetic properties of the surrounding accretion disk. In the framework of the steady-state Novikov-Thorne model, we calculated the radiation flux, the thermal spectra and luminosity of disk. For the first BH solution with free parameters $M$ and $a_{0}$, we showed that by increasing the compactness parameter both the Hernquist energy density and the tangential pressure (that mimics the dark matter's angular velocity in the static space-time) increase. Intuitively the increase of the tangential pressure leads to decrease the ISCO radius and we expect that the surface of disk increases and so the electromagnetic properties of the thin accretion disk increase in accordance with Tables ~\ref{TT1} and ~\ref{TT2}. For the second BH solution we also showed that with increasing the state parameter $\omega$, the tangential pressure increases and thus the ISCO radius decreases. The summarized results in Table ~\ref{TT3} show that similar to Cardoso's BH the electromagnetic properties of disk such as its luminosity increase. Comparing our results to the vacuum solutions without dark matter points out that in the presence of dark matter the ISCO radius decreases and so the energy flux, disk temperature and disk luminosity increase. Therefore, accretion disk luminosity for galactic BHs surrounded by dark matter is higher than the galactic BHs without dark matter. Also, accretion disk properties can distinguish between galaxies harboring supermassive BHs with and without dark matter fluid.

Finally, it is worth to note that although the BHs we studied here are exact general relativistic solutions of the field equations with more realistic density profiles, the qualitative study of the dark matter effects on the accretion disks for a  Schwarzschild BH immersed in a dark matter envelope has been done in Refs. \cite{dm1}--\cite{dm2}. Exploring luminosity of BH's accretion disk in these models also shows that the presence of dark matter increase the luminosity of disk in comparison to the pure Schwarzschild case. Although this makes it difficult to distinguish between various dark matter models, but the same results may allow such models to be distinguishable from other compact objects in alternative gravity theories or BH mimickers.

\section*{Acknowledgments}
The work of Mohaddese Heydari-Fard is supported by the Iran National Science Foundation (INSF) and the Research Council of Shahid Beheshti University under research project No. 4016024.

\end{multicols}
\end{document}